\documentclass[11pt]{article}

\usepackage[final]{acl}

\usepackage{times}
\usepackage{latexsym}
\usepackage[T1]{fontenc}
\usepackage[utf8]{inputenc}

\usepackage{microtype}
\usepackage{makecell}
\usepackage{inconsolata}

\usepackage{graphicx}
\usepackage{enumitem}
\usepackage{algorithm}
\usepackage{algorithmic}
\usepackage{array}       % 列格式优化
\usepackage{booktabs}    % 用于绘制专业三线表
\usepackage{xcolor}      % 用于灰色行高亮
\usepackage{amsmath}     % 用于上标排版
\usepackage{multirow}    % 合并单元格
\usepackage{amssymb} 
\usepackage[table]{xcolor}  % 必加table选项
\usepackage{graphicx}
\usepackage{subcaption}

\usepackage[table]{xcolor}
\newcommand{\retavg}[1]{\textcolor{green!60!black}{\textbf{#1}}}
\newcommand{\retwin}[1]{\textcolor{green!60!black}{\textbf{#1}}}
\newcommand{\effsub}[1]{_{\scriptstyle\,{\textcolor{green!60!black}{\mathbf{#1}}}}}
\title{DoPR: Reusable Compressed Document Prefixes for Efficient LLM Reranking}

\author{
  Beiya Dai$^{1}$,
  Yifan Wei$^{2}$,
  Guang Yang$^{2}$,
  Xing Shi$^{2}$,
  Xinbing Wang$^{3}$,
  Zhouhan Lin$^{1}$\thanks{
    Corresponding Author.
  } \\
  $^{1}$LUMIA Lab, Shanghai Jiao Tong University, Shanghai, China \\
  $^{2}$ByteDance, Beijing, China \\
  $^{3}$Shanghai Jiao Tong University, Shanghai, China \\
  \texttt{beiya\_dai@sjtu.edu.cn, lin.zhouhan@gmail.com}
}
\begin{document}
\maketitle
\begin{abstract}

% Large language models (LLMs) are highly effective rerankers, but pointwise reranking repeatedly processes the same document for different queries, causing substantial redundant document-side computation. We propose \textbf{DoPR}, a compressed document prefix framework that represents each document with query-independent prefix representations, decoupling offline document processing from online reranking. DoPR first compresses each document into a small set of reusable document prefixes, which are precomputed and stored offline. At inference time, these compressed document prefixes are reused across queries and injected as pre-filled prefix states, so the reranker only processes the query and scoring token online. This design reduces online cost through both document-side compression and cross-query prefix reuse. Experiments on TREC DL, BEIR, and BRIGHT with Qwen3 models from $0.6$B to $8$B show that DoPR substantially reduces memory footprint and inference latency while retaining competitive performance.

Large language models (LLMs) are effective rerankers, but pointwise reranking repeatedly processes the same document across different queries, causing substantial redundant document-side computation. We propose \textbf{DoPR}, a compressed document prefix framework that decouples offline document processing from online reranking. DoPR first selects query-independent document representations and converts them into compressed document prefix states, which are precomputed offline and reused whenever the document is retrieved. During online reranking, the model scores each query-document pair by processing only the query and scoring token, with document information supplied by the stored prefix states. This design reduces online cost through both document-side compression and cross-query prefix-state reuse. Experiments on TREC DL, BEIR, and BRIGHT with Qwen3 models from $0.6$B to $8$B show that DoPR achieves up to 8.0$\times$ online document-side memory reduction and up to 8.04$\times$ latency speedup, while retaining \textbf{97.1\%-99.5\%} of the average NDCG@10 of matched full-document rerankers.\footnote{Code available at: https://github.com/dbylynn/DoPR.}

\end{abstract}

\section{Introduction}
% Neural rerankers typically score each query-document pair by jointly encoding the query and document, enabling fine-grained interaction for relevance estimation~\citep{nogueira2019passage,nogueira2019multi,nogueira2020document,zhuang2023rankt5}. Recent LLM rerankers further strengthen this paradigm through prompting, pairwise/listwise ranking, and supervised adaptation~\citep{sun2023chatgpt,pradeep2023rankzephyr,qin2024large,yoon2024listt5,tang2024found}. While pointwise reranking is effective, it introduces substantial redundant computation. In practice, since the same document is often reranked for many different queries, the standard approach of re-encoding the full document for every query-document pair becomes prohibitively costly under strict latency constraints.
Pointwise reranking is a practical paradigm for retrieval pipelines, where each query-document pair is scored independently. This formulation is easy to parallelize, batch, and integrate with existing candidate-generation systems, while still allowing the model to capture fine-grained query-document interactions~\citep{nogueira2019passage,nogueira2019multi,nogueira2020document,zhuang2023rankt5,ma2024fine}. Its main drawback is redundant document-side computation. In real retrieval systems, the same document may be retrieved for many different queries over time, yet standard pointwise rerankers re-encode the full document for every query-document pair. This repeated processing becomes a major bottleneck for deploying LLM rerankers under strict latency and memory constraints.

% A natural way to reduce this redundancy is to shorten or reuse document-side computation during online reranking.
Existing efficiency methods mainly follow two directions: reducing the input processed in each reranking instance or precomputing document-side representations for reuse.
Prior work has reduced online cost through prompt compression, token pruning, or alternative reranking paradigms~\citep{mu2023learning,chevalier2023adapting,kim2022learned}. However, these methods mainly reduce the cost of each inference instance, and often still require reprocessing the same document when it is paired with different queries. Another line of work explores reusable document-side representations, such as succinct document representations, precomputed term representations, and late-interaction retrieval~\citep{cohen2022sdr,macavaney2020efficient,khattab2020colbert}. These methods precompute document information, but typically use it for retrieval-stage matching or specialized interaction functions rather than as internal states of an LLM reranker. This leaves open the question of whether document-side computation can be reused inside pointwise LLM reranking itself.

To address this question, we propose \textbf{DoPR}, a compressed document prefix framework that makes document-side computation reusable inside pointwise LLM reranking. The key idea is to represent each document with a small set of compressed document representations and convert them into compressed document prefix states, so that document-side computation can be performed once offline and reused across different queries. DoPR first selects salient document representations using self-attention signals and then converts them into compressed document prefix states. During online reranking, the stored prefix states are injected into the reranker, allowing the model to score each query-document pair by processing only the query and scoring token. In this way, DoPR reduces online cost by combining compact document prefixes with cross-query document reuse.

% We evaluate DoPR on standard reranking benchmarks using Qwen3 models ranging from 0.6B to 8B parameters~\citep{yang2025qwen3}. Experimental results show that DoPR achieves a strong effectiveness-efficiency trade-off, maintaining competitive ranking quality while substantially reducing token usage, memory footprint, and inference latency in the online reranking stage. Our efficiency analysis focuses on the setting where document prefixes are precomputed offline, matching practical retrieval systems in which documents are indexed once and reused across queries. We further observe that larger backbones generally better tolerate the prefix bottleneck on standard reranking benchmarks, while reasoning-intensive tasks show more task-level variation.
We evaluate DoPR on TREC DL, BEIR, and BRIGHT using Qwen3 backbones from 0.6B to 8B parameters~\citep{yang2025qwen3}. Our experiments are designed to examine whether reusable document prefixes can preserve the ranking quality of full-document pointwise rerankers while reducing online computation. Across standard and reasoning-intensive reranking benchmarks, DoPR retains competitive effectiveness relative to matched Qwen3 rerankers, with stronger retention on larger backbones in standard reranking settings. At the same time, DoPR substantially reduces online memory footprint and inference latency. We further analyze cross-query reuse and prefix-budget trade-offs, showing that the efficiency benefit is most pronounced when documents are long or repeatedly retrieved across queries. 
Overall, these results suggest that reusable document prefixes are especially suitable for retrieval systems with stable document collections and repeated document access.

% Our contributions are summarized as follows:
% \begin{itemize}[leftmargin=*]
%      \item We identify cross-query document-side reuse as an important efficiency problem in pointwise LLM reranking, where the same document is repeatedly processed online for different queries.
    
%     \item We propose \textbf{DoPR}, a compressed document prefix framework that makes document prefixes independent of the current query, enabling them to be precomputed offline and injected as pre-filled internal states during online reranking.

%     \item We provide a simple end-to-end instantiation using self-attention-based prefix selection, and show on TREC DL, BEIR, and BRIGHT that DoPR significantly reduces online memory and latency while retaining competitive performance.
% \end{itemize}
% \vspace{-5pt}
\section{Related Work}
\subsection{Compression for Efficient Reranking}

A common way to reduce reranking cost is to shorten the input processed online. Passage-embedding-based rerankers, such as PE-Rank~\citep{liu2025leveraging} and E$^2$Rank~\citep{liu2025e2rank}, replace passages or documents with compact embedding representations to reduce the effective input length for LLM reranking. More broadly, prompt and context compression methods improve LLM inference efficiency by learning compact prompt representations, compressing contexts into special tokens, or selecting informative context segments~\citep{mu2023learning,chevalier2023adapting,ge2023context,li2025prompt,bai2024beyond}. Token pruning and sparse attention methods further improve transformer efficiency by selecting salient tokens or reducing unnecessary attention computation~\citep{kim2022learned,wang2021spatten}. 
These methods reduce the cost of each inference instance, but their compressed inputs are usually tied to a specific query, prompt, or reranking context. This differs from our setting, where the key requirement is query-independent document compression for reuse across different queries.

\begin{figure*}[ht]
    \centering
    \includegraphics[width=\linewidth]{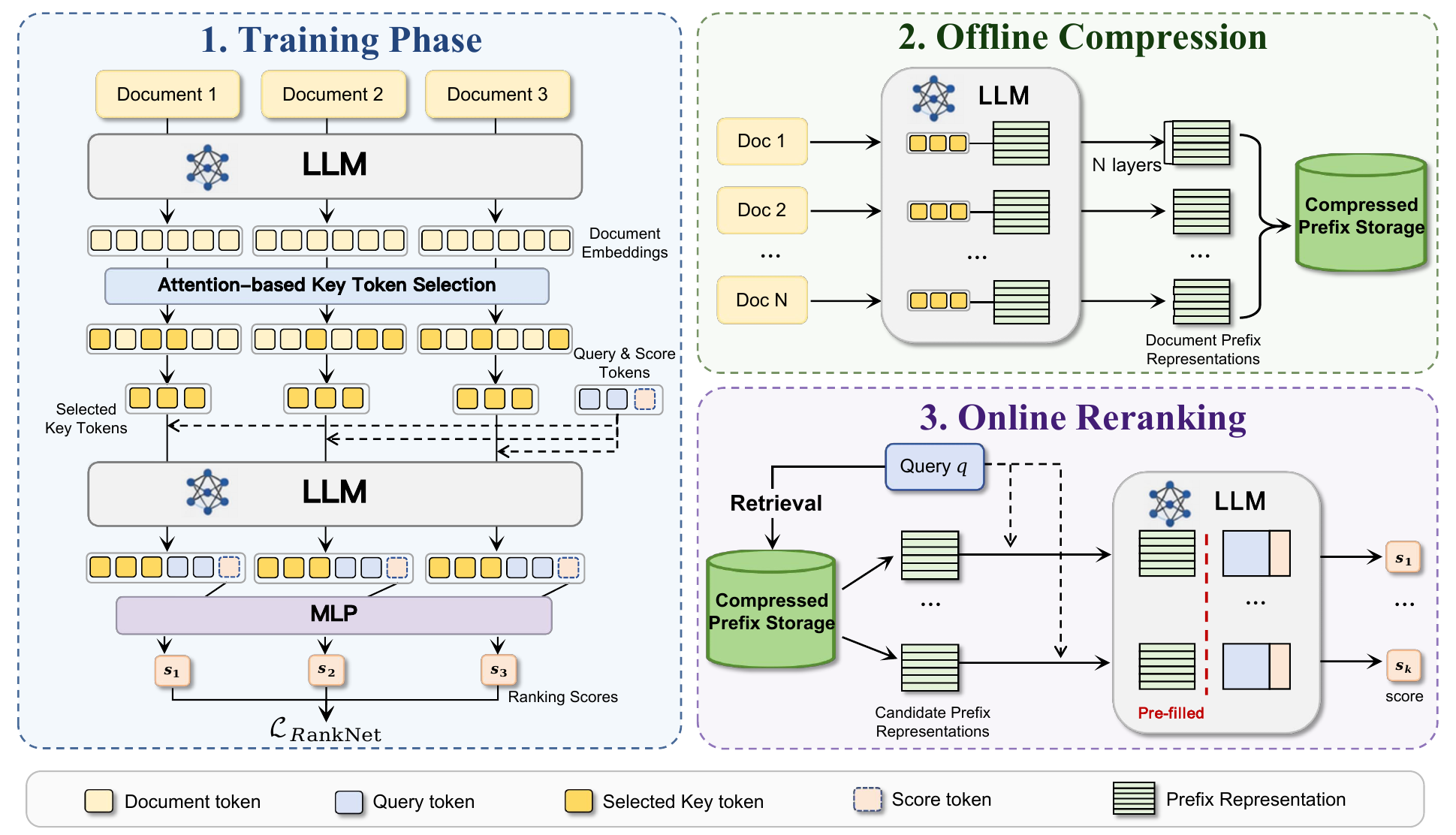}
    \caption{\textbf{Overview of DoPR.} The framework consists of three stages: training phase, offline compression, and online reranking. 
    During training, DoPR selects salient document token states as compressed document representations and optimizes the reranker end-to-end with the RankNet loss. During deployment, the selected document representations are converted into multi-layer prefix states, stored offline, and injected as pre-filled prefix states for efficient online reranking.
    }
    \label{fig:training}
    \vspace{-6pt}
\end{figure*}

\subsection{Reusable Document-side Representations}

Another relevant line of work precomputes document-side representations so that part of the document processing can be reused at query time. Succinct document representations compress documents into shorter forms for neural reranking~\citep{cohen2022sdr}, while precomputed term representations avoid repeatedly running transformer encoders over document terms~\citep{macavaney2020efficient}. Late-interaction retrieval models such as ColBERT precompute contextualized document token representations and perform efficient query-time interaction with the query~\citep{khattab2020colbert}. 
These methods share DoPR's motivation of reducing repeated document-side computation, but are typically designed for retrieval-specific architectures or earlier neural reranking pipelines rather than pointwise LLM reranking. 
DoPR instead reuses query-independent document prefix states inside a pointwise LLM reranker, avoiding repeated full-document processing during online reranking.
% These methods share DoPR's motivation of reducing repeated document-side computation. However, they are typically designed for retrieval-specific architectures or earlier neural reranking pipelines, rather than pointwise LLM reranking with pre-filled compressed document prefixes. DoPR differs by reusing query-independent document prefixes as internal states of a pointwise LLM reranker, avoiding repeated full-document processing during online reranking.
%These methods share DoPR's motivation of reducing repeated document-side computation. However, they are typically designed for retrieval-specific architectures or earlier neural reranking pipelines, rather than pointwise LLM reranking with pre-filled compressed document prefixes. In contrast, DoPR instead studies whether reusable document-side computation can be incorporated into pointwise LLM reranking itself. It stores query-independent compressed document prefixes offline and injects them as internal prefix states during online reranking, preserving the LLM reranker interface while avoiding repeated full-document processing.

\section{Method}
In this section, we present DoPR, a compressed document prefix framework for efficient pointwise LLM reranking, with three stages: training with a document-representation bottleneck, offline construction of document prefixes, and online reranking with stored prefix states, as shown in Figure~\ref{fig:training}.
% We present DoPR, a compressed document prefix framework for efficient pointwise LLM reranking, with three stages: prefix-bottleneck training, offline prefix construction, and online reranking with document prefixes, as shown in Figure~\ref{fig:training}.

\subsection{Problem Formulation}
\label{sec:problem_formulation}
Given a query $q$ and a candidate document $d = \{d_1, d_2, \dots, d_n\}$, a conventional pointwise reranker computes a relevance score by jointly encoding the full query-document pair:
\begin{equation}
    s(q,d) = f([q; d]),
\end{equation}
where $f(\cdot)$ denotes an LLM-based scoring function. This formulation allows full query-document interaction, but it also requires the document to be processed online for every query-document pair.

DoPR separates reusable document-side computation from query-time scoring. During training, the document is represented by a compact set of selected document representations:
\begin{equation}
    C_d = \Phi_{\text{train}}(d) = \{c_1,\dots,c_{K_{\text{train}}}\}, \quad K_{\text{train}} \ll n,
\end{equation}
where each $c_i \in \mathbb{R}^{d_h}$ is selected from document token states and depends only on $d$. At deployment time, the selected document representations are fed as prefix
inputs to the reranker, and the resulting multi-layer key-value states are
stored as compressed document prefix states:
\begin{equation}
    \widetilde{C}_d = \Phi_{\text{infer}}(d), 
    \quad K_{\text{infer}} \ll n.
\end{equation}

Here, $C_d$ denotes the selected document representations used to train the prefix bottleneck, while $\widetilde{C}_d$ denotes the compressed document prefix states obtained by forwarding these representations as prefix inputs through the reranker.
At inference time, $\widetilde{C}_d$ is injected as pre-filled prefix states, and the online input contains only $[q;t_{\text{score}}]$.

\subsection{Training Phase}
\label{sec:training_phase}

To make the selected document representations effective for ranking, DoPR trains document representation selection and relevance scoring in a unified end-to-end framework. During training, the model first identifies a small set of salient document token states as compressed document representations, then enforces a strict document-to-query bottleneck with a structured attention mask, and finally optimizes the resulting representations with a pairwise RankNet loss ~\citep{burges2005learning}.

\subsubsection{Attention-based Key Token Selection}
Given a document $d=\{d_1,d_2,\dots,d_n\}$, we first encode it independently using an LLM and obtain the final-layer token states:
\begin{equation}
    \mathbf{H}_d=[h_1,h_2,\dots,h_n] \in \mathbb{R}^{n \times d_h},
\end{equation}
where $d_h$ is the hidden size. Let $\bar{\mathbf{A}}_d \in \mathbb{R}^{n \times n}$ denote the final-layer self-attention matrix averaged over attention heads, where rows correspond to attending positions and columns correspond to attended positions.

We use attention concentration as a lightweight salience signal, based on the intuition that tokens strongly attended to by the document context can serve as compact carriers of document information. This avoids introducing an additional trainable selector. For token $d_i$, let $\mathcal{T}_p(\bar{\mathbf{A}}_{d,:,i})$ denote the set of the top-$p$ values in the $i$-th column of $\bar{\mathbf{A}}_d$. We define its salience score as
\begin{equation}
    u_i =
    \frac{1}{|\mathcal{T}_p(\bar{\mathbf{A}}_{d,:,i})|}
    \sum_{w \in \mathcal{T}_p(\bar{\mathbf{A}}_{d,:,i})} w,
\end{equation}
where $|\mathcal{T}_p(\bar{\mathbf{A}}_{d,:,i})|=\min(p,n-i+1)$ under causal attention. We then rank all document tokens according to $\mathbf{u}=\{u_1,\dots,u_n\}$ and select the indices of the top-$K_{\text{train}}$ tokens, denoted by $\mathcal{I}_{\text{top}}$. The corresponding hidden states are used as the salient token representations:
\begin{equation}
    C_d=\{h_i \mid i \in \mathcal{I}_{\text{top}}\}.
\end{equation}

The resulting $C_d \in \mathbb{R}^{K_{\text{train}} \times d_h}$ forms the selected document representations, which serve as the only document-side information path to the query during training.

\subsubsection{Structured Attention Mask}

\begin{figure}[htbp]
    \centering
    \includegraphics[width=0.92\linewidth]{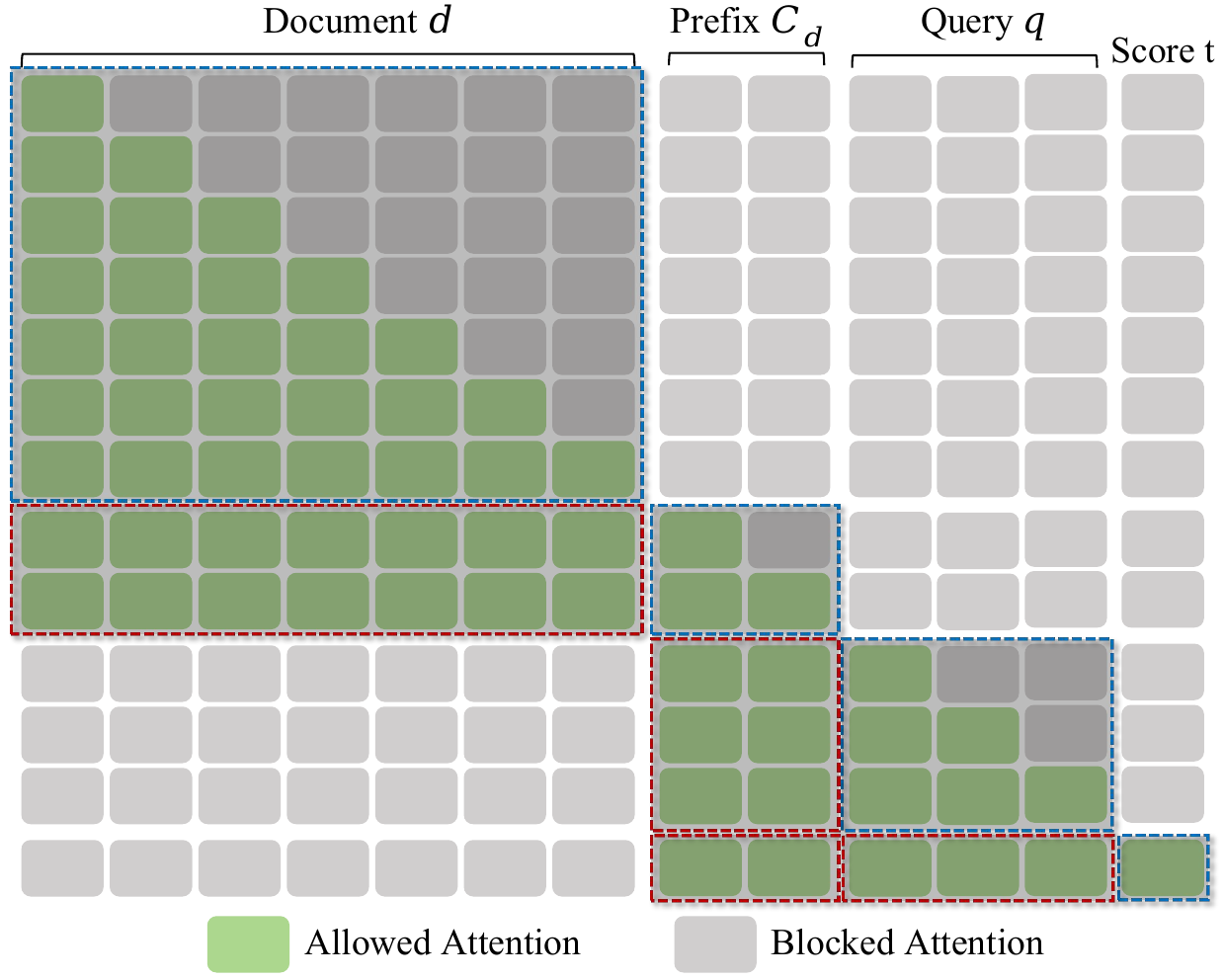}
    \caption{Structured attention mask used in training. 
     The query and score tokens access document information only through the selected document representations.}
    \label{fig:compress_attention}
    \vspace{-6pt}
\end{figure}

To align training with deployment, we impose a structured attention mask on the training input
\begin{equation}
    x = [d; C_d; q; t_{\text{score}}].
\end{equation}
As shown in Figure~\ref{fig:compress_attention}, the raw document tokens are encoded independently, and the query and score tokens can access document information only through $C_d$. This makes $C_d$ the sole document-to-query information path during training, encouraging the selected representations to carry ranking-relevant document information for the online setting where raw document tokens are unavailable.

% This design makes $C_d$ the only information path from the document to the query during training. It therefore forces the model to store ranking-relevant document information in the salient token representations, matching the online setting where raw document tokens are unavailable.

\subsubsection{Training Objective}

After the masked sequence is processed by the LLM, we take the hidden state of the score token, denoted by $h_{\text{score}}$, and feed it into a lightweight MLP head:
\begin{equation}
    s(q,d)=\mathrm{MLP}(h_{\text{score}}).
\end{equation}

Although each forward pass yields a pointwise relevance score, the model is trained using the pairwise RankNet objective. Given a query $q$ and a pair of candidate documents $(d^+,d^-)$, we compute
\begin{equation}
    P(d^+ \succ d^- \mid q)=\sigma\big(s(q,d^+) - s(q,d^-)\big),
\end{equation}
where $\sigma(\cdot)$ denotes the sigmoid function. The loss is
\begin{equation}
\begin{split}
\mathcal{L}_{\text{RankNet}}
&= -\log P(d^{+} \succ d^{-} \mid q) \\
&= \log \bigl(1 + \exp\bigl(-(s(q,d^{+}) - s(q,d^{-}))\bigr)\bigr).
\end{split}
\end{equation}

Since the loss is backpropagated through the score token, query-representation interactions, and selected document token states, the selected document representations are optimized for ranking rather than generic reconstruction.
\begin{table*}[htbp]
    \centering
    \renewcommand{\arraystretch}{1.1}
    \resizebox{\linewidth}{!}{
    \begin{tabular}{lcc|ccccccccc}
    \toprule
    \textbf{Model} 
    & \textbf{DL19} 
    & \textbf{DL20} 
    & \textbf{Covid} 
    & \textbf{NFCorpus} 
    & \textbf{Touche} 
    & \textbf{DBPedia} 
    & \textbf{SciFact} 
    & \textbf{Signal} 
    & \textbf{News} 
    & \textbf{Robust} 
    & \textbf{Avg.} \\
    \midrule

    BM25 
    & 50.58 & 47.96 & 59.47 & 30.75 & 44.22 & 31.80 & 67.89 & 33.05 & 39.52 & 40.70 & 43.43 \\

    \midrule

    MonoBERT-340M 
    & 70.50 & 67.28 & 70.01 & 36.88 & 31.75 & 41.87 & 71.36 & 31.44 & 44.62 & 49.35 & 47.16 \\

    MonoT5-3B 
    & 71.83 & 68.89 & 80.71 & 37.30 & 32.20 & 48.30 & 58.50 & 76.30 & 32.50 & 44.80 & 51.33 \\

    RankT5-3B 
    & 72.95 & 69.63 & 82.00 & 37.40 & 31.90 & 49.50 & 58.30 & 77.10 & 38.08 & 45.00 & 52.50 \\
    
    RankLLaMA-7B & 73.28 & 68.55 & 80.15 & 32.25 & 35.79 & 44.50 &  70.87 & 28.60 & 42.01 &  35.74 & 46.23\\

    RankZephyr 
    & 73.39 & 70.02 & 83.20 & 37.60 & 32.40 & 44.50 & 74.90 & 31.50 & 52.50 & 54.30 & 51.36 \\

    RankGPT-4o 
    & 74.78 & 69.52 & 83.41 & 39.67 & 32.26 & 45.56 & 77.41 & 34.20 & 51.92 & 60.25 & 53.09 \\

    E$^2$Rank-8B
    & 72.95 & 71.16 & 84.09 & 39.08 & 42.06 & 43.44 & 77.49 & 34.01 & 54.25 & 60.34 & 54.35 \\

    \midrule

    Qwen3-Rerank-0.6B 
    & 75.30 & 69.94 & 85.19 & 39.13 & 41.18 & 45.14 & 76.99 & 33.73 & 52.51 & 63.50 & 54.67 \\

    \rowcolor[HTML]{F2F3F5}
    \textbf{DoPR-0.6B} 
    & 74.30 & 68.58 & 84.78 & 38.20 & 37.86 & 44.26 & 76.31 & 31.78 & 51.96 & 59.56 & 53.09 \\

    \rowcolor[HTML]{EAF3FF}
    \textit{Retention} 
    & 98.7\% & 98.1\% & 99.5\% & 97.6\% & 91.9\% & 98.1\% & 99.1\% & 94.2\% & 99.0\% & 93.8\% & \retavg{97.1\%} \\

    \midrule

    Qwen3-Rerank-4B 
    & 75.93 & 70.95 & 85.86 & 39.76 & 37.93 & 45.99 & 78.82 & 32.89 & 53.13 & 66.62 & 55.13 \\

    \rowcolor[HTML]{F2F3F5}
    \textbf{DoPR-4B} 
    & 75.35 & 71.19 & 85.25 & 39.53 & 37.06 & 45.44 & 78.62 & 33.52 & 52.67 & 65.90 & 54.75 \\

    \rowcolor[HTML]{EAF3FF}
    \textit{Retention} 
    & 99.2\% & \retwin{100.3\%} & 99.3\% & 99.4\% & 97.7\% & 98.8\% & 99.7\% & \retwin{101.9\%} & 99.1\% & 98.9\% & \retavg{99.3\%} \\

    \midrule

    Qwen3-Rerank-8B 
    & 76.86 & 71.57 & 85.30 & 40.38 & 37.88 & 46.74 & 80.02 & 32.57 & 52.86 & 67.09 & 55.36 \\

    \rowcolor[HTML]{F2F3F5}
    \textbf{DoPR-8B} 
    & 76.17 & 70.44 & 85.42 & 40.68 & 37.14 & 45.38 & 80.27 & 32.90 & 52.64 & 66.05 & 55.06 \\

    \rowcolor[HTML]{EAF3FF}
    \textit{Retention} 
    & 99.1\% & 98.4\% & \retwin{100.1\%} & \retwin{100.7\%} & 98.0\% & 97.1\% & \retwin{100.3\%} & \retwin{101.0\%} & 99.6\% & 98.4\% & \retavg{99.5\%} \\

    \bottomrule
    \end{tabular}}
    \caption{
    Performance on TREC DL and BEIR.
    We compare DoPR with Qwen3-Rerank across three model scales, together with representative rerankers.
    \textit{Retention} denotes the NDCG@10 retained by DoPR relative to the Qwen3-Rerank.
    % DoPR retains \retavg{97.1\%-99.5\%} of average full-document effectiveness while replacing online full-document processing with compressed prefixes that are precomputed offline and reused across queries.
    }
    \vspace{-6pt}
    \label{tab:dl_beir_performance}
\end{table*}
\subsection{Offline Compression}
\label{sec:offline_compression}

At deployment time, we process the document collection offline. For each document $d$, we first run a document-only forward pass to identify the top-$K_{\text{infer}}$ salient document positions and extract their representations. These selected representations are then used as document prefix inputs to the reranker, and the resulting multi-layer key-value states are stored as compressed document prefix states.

Since $\widetilde{C}_d$ is query-independent, it can be reused whenever the same document is retrieved. The inference prefix budget $K_{\text{infer}}$ can be set independently of $K_{\text{train}}$, allowing deployment-time trade-offs between ranking quality and online cost.

\subsection{Online Reranking}
\label{sec:online_reranking}
At inference time, we retrieve $\widetilde{C}_d$ for each candidate document and inject it into the reranker as pre-filled prefix states.~\footnote{The stored prefix states are not the key-value cache of a document-only forward pass. They are the multi-layer key-value states obtained by forwarding the selected document representations as prefix inputs to the reranker.
} The model then processes only the query and score token online, while document-side information is provided by the stored prefix states. Compared with full-document reranking, DoPR reduces the online document-side budget from $n$ document tokens to $K_{\text{infer}}\ll n$ stored prefix states, with efficiency gains coming from both document-side compression and cross-query prefix state reuse.

\section{Experiments}
    \subsection{Implementation Details}
     We implement \textbf{DoPR} using models from the Qwen3~\citep{yang2025qwen3} series as the backbone, with parameter sizes ranging from 0.6B to 8B. We adopt the training dataset introduced in \citep{liu2025e2rank}. All models are trained using AdamW with a learning rate of $5\times10^{-6}$ and a total batch size of 32 on 8 NVIDIA A100 GPUs. The ranking head is optimized using the RankNet loss. Unless otherwise noted, we use a default training prefix budget of $K_{\text{train}}=32$. At inference time, $K_{\text{infer}}$ is set separately for each benchmark according to its document length budget, and the full benchmark-specific settings are summarized in Appendix~\ref{sec:compression_settings}.
    
    % \vspace{-6pt}
    \subsection{Datasets and Metrics}
    We evaluate DoPR on three widely used reranking benchmarks covering both in-domain and out-of-domain retrieval scenarios. First, we use the TREC DL19 and DL20 test sets~\citep{craswell2025overview}, which are standard benchmarks for passage reranking with dense human relevance judgments. Second, we evaluate on BEIR~\citep{thakur2021beir} to assess out-of-domain generalization across diverse retrieval domains. Following prior work, we report results on representative subsets covering scientific, financial, argumentative, and fact verification scenarios. Third, we evaluate on BRIGHT~\citep{su2024bright}, a challenging benchmark for reasoning-intensive retrieval, to test whether compressed document prefixes preserve reasoning-relevant information. We use \emph{NDCG@10} as the primary evaluation metric across all datasets.
     
    \begin{table*}[htbp]
    \centering
    \renewcommand{\arraystretch}{1.05}
    \resizebox{\linewidth}{!}{
    \begin{tabular}{lccccccccccccc}
    \toprule
    \multirow{2}{*}{\textbf{Model}} 
    & \multicolumn{7}{c}{\textbf{StackExchange}} 
    & \multicolumn{2}{c}{\textbf{Coding}} 
    & \multicolumn{3}{c}{\textbf{Theorem-based}} 
    & \multirow{2}{*}{\textbf{Avg.}} \\
    \cmidrule(lr){2-8} \cmidrule(lr){9-10} \cmidrule(lr){11-13}
    & \textbf{Bio.} 
    & \textbf{Econ.} 
    & \textbf{Earth.} 
    & \textbf{Psy.} 
    & \textbf{Rob.} 
    & \textbf{Stack.} 
    & \textbf{Sus.} 
    & \textbf{Pony.} 
    & \textbf{LC.} 
    & \textbf{AoPS} 
    & \textbf{TheoQ.} 
    & \textbf{ThoT.} 
    & \\
    \midrule

    ReasonIR 
    & 43.5 & 32.8 & 43.0 & 38.9 & 21.1 & 30.6 & 27.3 & 31.6 & 19.6 & 7.3 & 36.7 & 34.1 & 30.5 \\

    \midrule

    RankT5-3B 
    & 11.4 & 22.1 & 10.9 & 13.6 & 11.4 & 11.4 & 16.0 & 27.5 & 38.1 & 9.2 & 18.3 & 9.5 & 16.6 \\

    RankZephyr 
    & 19.9 & 17.4 & 12.4 & 34.9 & 24.7 & 13.4 & 22.3 & 29.3 & 32.4 & 6.1 & 29.0 & 30.1 & 22.6 \\

    ERank-4B 
    & 42.1 & 42.5 & 26.3 & 36.4 & 20.8 & 27.3 & 33.2 & 31.7 & 21.8 & 10.9 & 32.8 & 40.6 & 30.5 \\

    E$^2$Rank-8B 
    & 49.2 & 47.2 & 32.3 & 44.7 & 28.2 & 32.9 & 38.4 & 10.6 & 36.2 & 8.2 & 38.2 & 33.4 & 33.4 \\

    \midrule

    Qwen3-Rerank-0.6B 
    & 37.4 & 24.0 & 40.7 & 40.7 & 18.2 & 22.5 & 25.3 & 25.6 & 32.6 & 6.3 & 34.0 & 35.1 & 28.5 \\

    \rowcolor[HTML]{F2F3F5}
    \textbf{DoPR-0.6B} 
    & 34.1 & 25.8 & 38.1 & 37.0 & 18.2 & 25.0 & 26.4 & 29.1 & 31.5 & 9.3 & 32.1 & 30.6 & 28.1 \\

    \rowcolor[HTML]{EAF3FF}
    \textit{Retention} 
    & 91.2\% 
    & \retwin{107.5\%} 
    & 93.6\% 
    & 90.9\% 
    & 100.0\% 
    & \retwin{111.1\%} 
    & \retwin{104.3\%} 
    & \retwin{113.7\%} 
    & 96.6\% 
    & \retwin{147.6\%} 
    & 94.4\% 
    & 87.2\% 
    & \retavg{98.6\%} \\

    \midrule

    Qwen3-Rerank-4B 
    & 47.8 & 30.3 & 47.4 & 48.3 & 25.0 & 26.0 & 35.1 & 31.4 & 33.2 & 6.8 & 37.9 & 38.9 & 34.0 \\

    \rowcolor[HTML]{F2F3F5}
    \textbf{DoPR-4B} 
    & 45.1 & 29.0 & 47.0 & 47.7 & 25.0 & 28.8 & 34.8 & 36.3 & 26.5 & 7.1 & 37.8 & 38.7 & 33.7 \\

    \rowcolor[HTML]{EAF3FF}
    \textit{Retention} 
    & 94.4\% 
    & 95.7\% 
    & 99.2\% 
    & 98.8\% 
    & 100.0\% 
    & \retwin{110.8\%} 
    & 99.1\% 
    & \retwin{115.6\%} 
    & 79.8\% 
    & \retwin{104.4\%} 
    & 99.7\% 
    & 99.5\% 
    & \retavg{99.1\%} \\

    \midrule

    Qwen3-Rerank-8B 
    & 48.4 & 30.7 & 47.0 & 49.5 & 27.4 & 27.3 & 36.9 & 30.2 & 29.4 & 8.7 & 39.7 & 40.5 & 34.6 \\

    \rowcolor[HTML]{F2F3F5}
    \textbf{DoPR-8B} 
    & 46.9 & 30.0 & 44.7 & 50.7 & 27.9 & 28.2 & 36.1 & 27.8 & 30.6 & 8.1 & 38.7 & 40.9 & 34.2 \\

    \rowcolor[HTML]{EAF3FF}
    \textit{Retention} 
    & 96.9\% 
    & 97.7\% 
    & 95.1\% 
    & \retwin{102.4\%} 
    & \retwin{101.8\%} 
    & \retwin{103.3\%} 
    & 97.8\% 
    & 92.1\% 
    & \retwin{104.1\%} 
    & 93.1\% 
    & 97.5\% 
    & \retwin{101.0\%} 
    & \retavg{98.8\%} \\

    \bottomrule
    \end{tabular}
    }
    \caption{
    Performance on BRIGHT.
    We compare DoPR with Qwen3-Rerank across three model scales, together with representative rerankers.
    % \textit{Retention} denotes the NDCG@10 retained by DoPR relative to the Qwen3 reranker.
    % DoPR retains \retavg{98.6\%-99.1\%} of average full-document effectiveness under reasoning-intensive retrieval settings while replacing online full-document processing with compressed prefixes.
    }
    \label{tab:bright_performance}
    \vspace{-6pt}
\end{table*}
    
    \subsection{Baselines}
    Our primary baselines are the matched full-document Qwen3 rerankers at three model scales, denoted as \textbf{Qwen3-Rerank-0.6B}, \textbf{Qwen3-Rerank-4B}, and \textbf{Qwen3-Rerank-8B}. These models use the same backbone architecture, training data, scoring head, and optimization setup as DoPR, but process the full query-document pair without compression. This comparison directly measures how much ranking effectiveness is retained when full online document processing is replaced with query-independent compressed prefixes. 

    \vspace{-8pt}
    We also compare with representative reranking methods. On TREC DL and BEIR, we include BM25~\citep{robertson2009probabilistic} as a lexical retrieval baseline, and MonoBERT~\citep{nogueira2019passage} and MonoT5~\citep{nogueira2020document} as classic supervised neural rerankers. We further include fine-tuned rerankers with different backbones and ranking paradigms, including RankT5~\citep{zhuang2023rankt5}, RankLLaMA~\citep{ma2024fine}, and RankZephyr~\citep{pradeep2023rankzephyr}, covering T5-based supervised reranking, LLM-based pointwise reranking, and LLM-based listwise reranking, respectively. In addition, we report RankGPT-4o as a prompting-based LLM reranker and E$^2$Rank~\citep{liu2025e2rank} as a recent efficient reranking method. On BRIGHT, we report ReasonIR~\citep{shao2025reasonir} as a strong reasoning-oriented retrieval baseline, together with RankT5, RankZephyr, ERank~\citep{cai2026erankw}, and E$^2$Rank as representative neural, LLM-based, and efficient reranking references.

    \subsection{Main Results}
    We evaluate whether compressed prefixes can retain the effectiveness of full-document reranking. Results are reported on TREC DL and BEIR for standard reranking, and on BRIGHT for reasoning-intensive retrieval.
    
    \subsubsection{Results on TREC DL and BEIR}
    On TREC DL and BEIR, DoPR retains most of the effectiveness of the corresponding Qwen3-Rerank baselines, as shown in Table~\ref{tab:dl_beir_performance}. 
    The average retention increases from 97.1\% at the 0.6B scale to 99.3\% and 99.5\% at the 4B and 8B scales, respectively, indicating that larger backbones better tolerate the prefix bottleneck. DoPR-0.6B shows larger drops on Touche, Signal, and Robust, while remaining close to the full-document baseline on most other datasets. At 4B and 8B, the effectiveness gap narrows substantially, and DoPR slightly outperforms the full-document baseline on several datasets. These results suggest that the stored prefix states preserve most ranking-relevant document information, and smaller models are more sensitive to information loss under a limited prefix budget.

    \begin{table*}[htbp] 
    \centering 
    \renewcommand{\arraystretch}{1.0} 
    \resizebox{\linewidth}{!}{
    \begin{tabular}{lccllccll} 
    \toprule 
    \multirow{2}{*}{\textbf{Model}} 
    & \multicolumn{4}{c}{\textbf{DL19}} 
    & \multicolumn{4}{c}{\textbf{Covid}} \\ 
    \cmidrule(lr){2-5} \cmidrule(lr){6-9} 
    & \textbf{\makecell{Doc \\ Budget}} 
    & \textbf{NDCG@10$\uparrow$} 
    & \textbf{Memory $\downarrow$} 
    & \textbf{Latency $\downarrow$} 
    & \textbf{\makecell{Doc \\ Budget}} 
    & \textbf{NDCG@10$\uparrow$} 
    & \textbf{Memory $\downarrow$} 
    & \textbf{Latency $\downarrow$} \\ 
    \midrule 

    Qwen3-Rerank-0.6B 
    & 256 & 75.30 & 28.00 & 17.612 
    & 1024 & 85.19 & 112.00 & 41.007 \\ 

    \rowcolor[HTML]{F2F3F5}
    \textbf{DoPR-0.6B} 
    & 32 & 74.30 
    & $3.50\effsub{8.0\times}$ 
    & $14.600\effsub{1.21\times}$ 
    & 128 & 84.78 
    & $14.00\effsub{8.0\times}$ 
    & $14.372\effsub{2.85\times}$ \\ 

    \midrule 

    Qwen3-Rerank-4B 
    & 256 & 75.93 & 36.00 & 29.386 
    & 1024 & 85.86 & 144.00 & 121.194 \\ 

    \rowcolor[HTML]{F2F3F5}
    \textbf{DoPR-4B} 
    & 32 & 75.35 
    & $4.50\effsub{8.0\times}$ 
    & $18.430\effsub{1.59\times}$ 
    & 128 & 85.25 
    & $18.00\effsub{8.0\times}$ 
    & $18.095\effsub{6.70\times}$ \\ 

    \midrule 

    Qwen3-Rerank-8B 
    & 256 & 76.86 & 36.00 & 37.756 
    & 1024 & 85.30 & 144.00 & 150.708 \\ 

    \rowcolor[HTML]{F2F3F5}
    \textbf{DoPR-8B} 
    & 32 & 76.17 
    & $4.50\effsub{8.0\times}$ 
    & $20.547\effsub{1.84\times}$ 
    & 128 & 85.42 
    & $18.00\effsub{8.0\times}$ 
    & $18.743\effsub{8.04\times}$ \\ 

    \bottomrule 
    \end{tabular}}
    \caption{
    Online efficiency comparison on DL19 and Covid. Doc Budget denotes the online document length for full-document rerankers and the retained prefix length for DoPR. Memory and latency are averaged per query-document pair during online reranking. 
    % Green numbers denote memory reduction ratios and latency speedups over the Qwen3-Rerank baseline.
    }
    \label{tab:efficiency_all} 
    \vspace{-6pt}
    \end{table*}
    \subsubsection{Results on BRIGHT}
    BRIGHT provides a more challenging test because many tasks require deeper semantic matching and reasoning. As reported in Table~\ref{tab:bright_performance}, DoPR retains 98.6\%, 99.1\%, and 98.8\% of the average NDCG@10 of Qwen3-Rerank at the 0.6B, 4B, and 8B scales, respectively. Unlike TREC DL and BEIR, the retention on BRIGHT is not strictly monotonic with model size, indicating that reasoning-intensive retrieval introduces stronger task-level variation. DoPR matches or exceeds the full-document baseline on several subsets, such as StackOverflow and Pony, but shows drops on subsets such as Biology and LC at certain model scales. This pattern suggests that stored prefix states can preserve much of the document information needed for semantic matching, but tasks requiring more dispersed evidence or fine-grained document-level matching may be more sensitive to a fixed prefix budget.

    \subsection{Efficiency Analysis}
    We analyze DoPR from three perspectives: online reranking cost, amortized benefit under document reuse, and effectiveness-efficiency trade-offs under different deployment settings.

    \subsubsection{Online Reranking Efficiency}
    Table~\ref{tab:efficiency_all} reports online reranking cost after compressed document prefix states have been precomputed offline. 
    DoPR reduces the document-side memory footprint by the same ratio as the prefix compression rate, while latency gains vary with document length and model scale. 
    % On DL19, where the full-document budget is relatively short, DoPR achieves moderate speedups of 1.21$\times$-1.84$\times$. On Covid, where documents are longer, the speedups become much larger, reaching 2.85$\times$-8.04$\times$. 
    DoPR achieves 1.21$\times$-1.84$\times$ speedups on DL19 and larger speedups of 2.85$\times$-8.04$\times$ on Covid, where documents are longer. 
    This pattern shows that DoPR is most beneficial when full-document reranking spends a larger fraction of online computation on document-side processing. Detailed efficiency results on additional datasets are provided in Appendix~\ref{sec:datasets_efficiency}.

    \subsubsection{Controlled Cross-query Reuse Analysis}
    Since DL19 and Covid do not contain repeated cross-query document reuse, we construct a controlled workload where the same document is scored for $r$ different queries. Let $T_{\text{off}}$ denote the one-time prefix construction cost, $T_{\text{full}}$ the full-document reranking latency, and $T_{\text{online}}$ the DoPR online latency. The total costs of full-document reranking and DoPR are defined as
    \begin{equation}
    \begin{aligned}
        T_{\text{Full}}(r) &= rT_{\text{full}}, \\
        T_{\text{DoPR}}(r) &= T_{\text{off}} + rT_{\text{online}}.
    \end{aligned}
    \end{equation}
    
    The amortized speedup is then
    \begin{equation}
    \mathrm{Speedup}(r)
    =
    \frac{T_{\text{Full}}(r)}{T_{\text{DoPR}}(r)}
    =
    \frac{rT_{\text{full}}}{T_{\text{off}}+rT_{\text{online}}}.
    \end{equation}

\begin{figure}
        \centering
        \includegraphics[width=\linewidth]{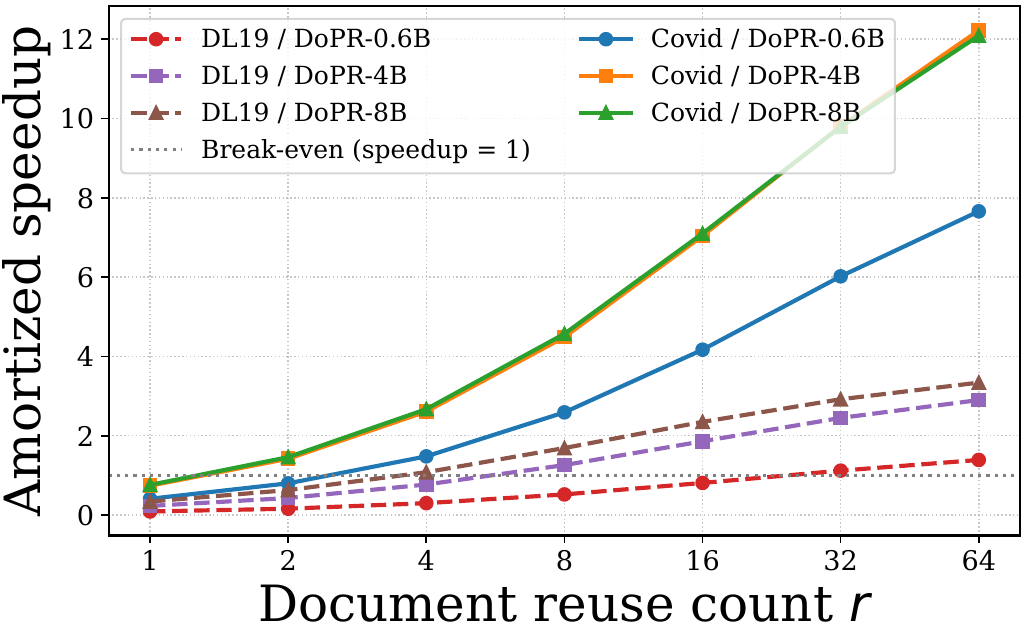}
        \caption{
        Amortized speedup under controlled document reuse. 
        % The same document is scored for $r$ different queries, and the reported speedup includes the one-time offline prefix construction cost of DoPR.
        }
        \label{fig:reuse_speedup}
        \vspace{-6pt}
    \end{figure}

    Figure~\ref{fig:reuse_speedup} shows that amortized speedup increases with the reuse count $r$, as the one-time offline construction cost is shared across more queries. 
    The benefit is more pronounced on Covid than on DL19 because longer documents make repeated full-document reranking more expensive, so the offline cost is amortized more quickly. 
    \footnote{DoPR-4B and DoPR-8B show similar trends on Covid, since their offline construction cost, full-document latency, and DoPR online latency scale similarly in this setting.}
    
       \begin{figure*}[ht]
    \centering
      \begin{subfigure}{0.245\textwidth}
        \includegraphics[width=\linewidth]{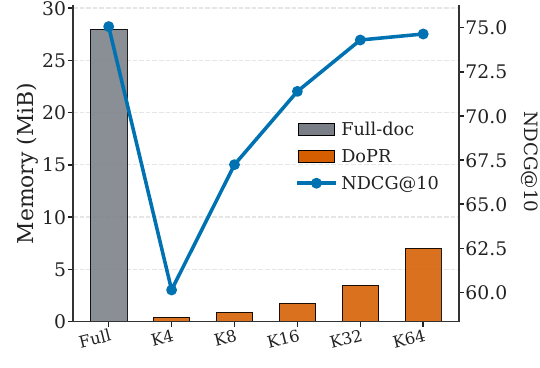}
        \subcaption{DL 19 Memory}
        \end{subfigure}
      % \hspace{8pt}
      \begin{subfigure}{0.245\textwidth}
        \includegraphics[width=\linewidth]{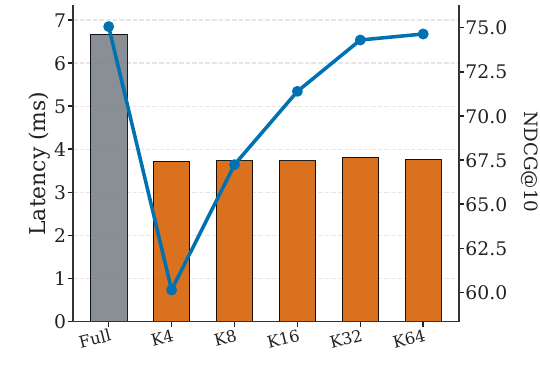}
        \subcaption{DL 19 Latency}
    \end{subfigure}
    \begin{subfigure}{0.245\textwidth}
        \includegraphics[width=\linewidth]
        {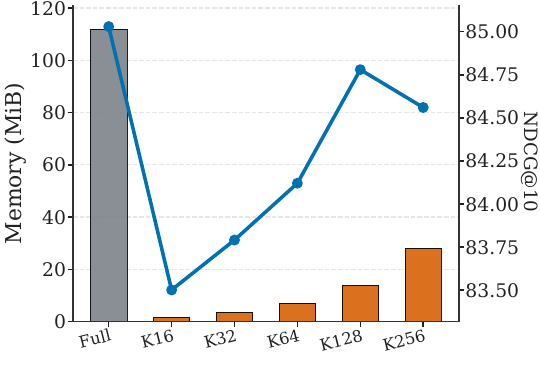}
        \subcaption{Covid Memory}
    \end{subfigure}
    \begin{subfigure}{0.245\textwidth}
        \includegraphics[width=\linewidth]{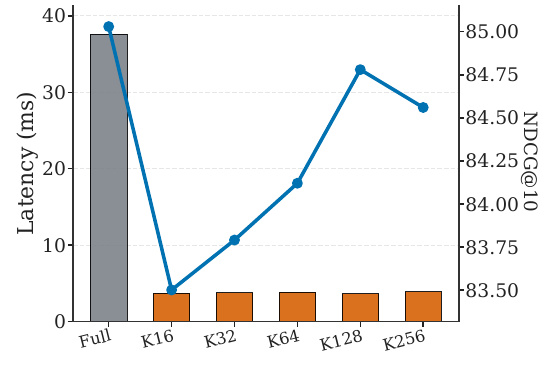}
        \subcaption{Covid Latency}
    \end{subfigure}
    \caption{
    Effectiveness-efficiency trade-off under different inference prefix budgets $K_{\text{infer}}$ on DL19 and Covid.
    We vary $K_{\text{infer}}$ and report the corresponding document-side memory footprint, reranking latency, and NDCG@10.
    }
    \label{fig:tradeoff_prefix_budget}
    \vspace{-6pt}
    \end{figure*}
    \subsubsection{Effectiveness and Efficiency Trade-off}
    A practical advantage of DoPR is that the inference prefix budget $K_{\text{infer}}$ can be adjusted without retraining. Figure~\ref{fig:tradeoff_prefix_budget} examines this trade-off using a single model trained with $K_{\text{train}}=32$. Increasing $K_{\text{infer}}$ allows the model to retain more document-side information and generally improves ranking performance, but it also increases memory footprint and latency. 
    % This gives DoPR flexible deployment operating points: smaller prefix budgets prioritize efficiency, while larger budgets recover more full-document effectiveness. 

\begin{figure}
        \centering
        \includegraphics[width=\linewidth]{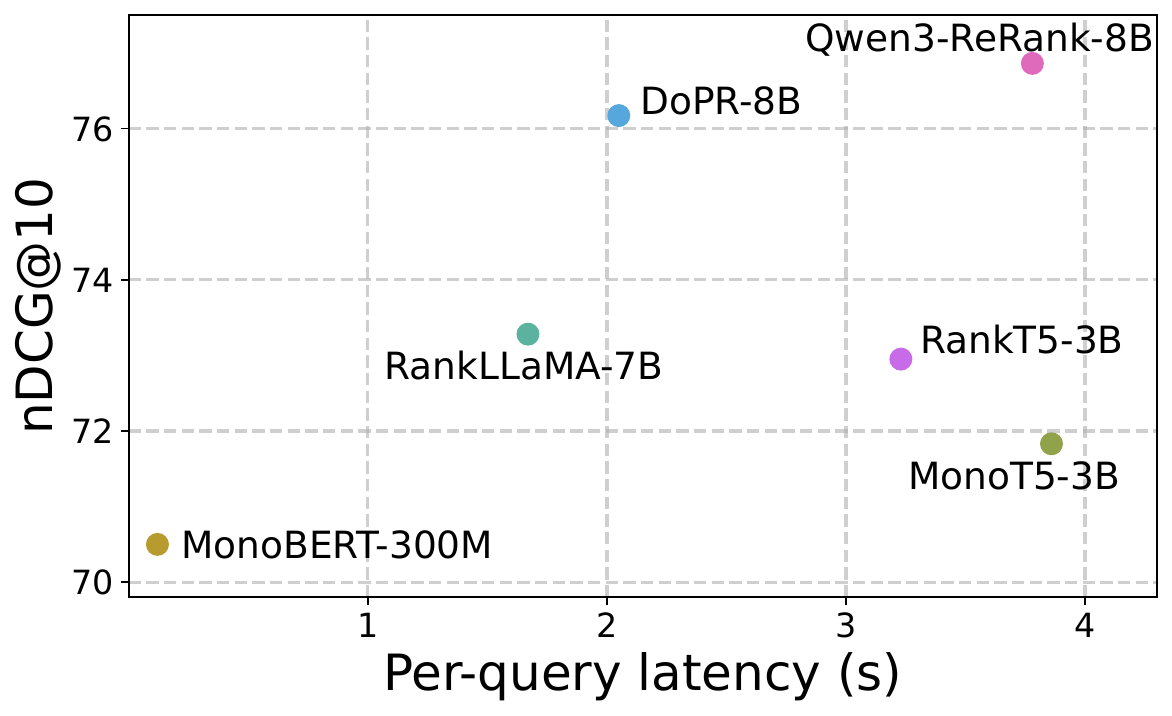}
        \caption{Effectiveness and efficiency trade-off on DL19. Up and left is better.}
        \label{fig:efficiency_comparison}
        \vspace{-6pt}
    \end{figure}

    We further compare DoPR with representative rerankers in Figure~\ref{fig:efficiency_comparison}. 
    DoPR-8B approaches the effectiveness of Qwen3-Rerank-8B with substantially lower per-query latency, while outperforming the remaining methods in nDCG@10. 
    We exclude listwise rerankers because their latency depends on joint candidate processing, decoding length, and batching strategy, making it not directly comparable to pointwise reranking.
    
    \begin{table}[htbp]
    \centering
    \resizebox{\linewidth}{!}{
    \begin{tabular}{lcccc}
    \toprule
    \textbf{Model} & \textbf{$K_{\text{train}}$} & \textbf{DL19} & \textbf{DL20} & \textbf{Avg.} \\
    \midrule
    Qwen3-Rerank & - & 75.30 & \textbf{69.94} & \textbf{72.62} \\ \hline
    % \multirow{5}{*}{DoPR} & 128 & 74.68 & 70.15 & 72.42 \\
    \multirow{5}{*}{DoPR} & 64 & \textbf{75.48} & 69.47 & 72.48 \\
    % \rowcolor[HTML]{F2F3F5}
    &32 & 74.30 & 68.58 & 71.44 \\
    &16 & 74.13 & 68.19 & 71.16 \\
    &8  & 72.83 & 66.45 & 69.64 \\
    &4  & 68.39 & 60.36 & 64.38 \\
    \bottomrule
    \end{tabular}}
    \caption{
    Effect of the number of compressed document prefixes $K_{\text{train}}$. 
    }
    \vspace{-6pt}
    \label{tab:compression_ratio}
    \end{table}
    \subsection{Ablation Studies}
    We conduct ablations to examine three key design choices in DoPR: the training prefix $K_{\text{train}}$, compression-free warmup, and the prefix construction strategy.
    
    \subsubsection{Effect of the Training Prefix $K_{\text{train}}$}
    \label{sec:training_prefix}
    We study the training prefix $K_{\text{train}}$, which controls the strength of the document-side bottleneck. We train separate models with $K_{\text{train}} \in \{4,8,16,32,64\}$ and evaluate them with matched inference budgets. As shown in Table~\ref{tab:compression_ratio}, larger budgets generally improve reranking effectiveness, while very small budgets impose an overly restrictive bottleneck and substantially decrease ranking performance. For our experiments, we use $K_{\text{train}}=32$ as the default setting. The corresponding training loss curves for different $K_{\text{train}}$ are provided in Appendix~\ref{sec:app-training-loss-vary-k}.

    \subsubsection{Effect of Compression-free Warmup}
    \label{sec:training_warmup}
    
    We next examine compression-free warmup, where the model is first trained as a full-document reranker and the prefix bottleneck is activated afterward. The goal is to stabilize the ranking behavior of the backbone before requiring the model to rely on compact prefix representations. As shown in Table~\ref{tab:warmup_steps}, compression-free warmup improves performance compared with training with compressed prefixes from the beginning. A moderate warmup period gives the best overall result, while longer warmup does not yield consistent additional gains. We therefore use 200 warmup steps by default.
      \begin{table}[htbp]
    \centering
    \resizebox{0.9\linewidth}{!}{
    \begin{tabular}{lccc}
    \toprule
    \textbf{Warmup Steps}  & \textbf{DL19} & \textbf{DL20}& \textbf{Avg.} \\
    \midrule
    0    & 73.76 & 66.79 & 70.28 \\
    \rowcolor[HTML]{F2F3F5}
    \textbf{200}  & 74.30 & \textbf{68.58} & \textbf{71.44} \\
    400  & \textbf{74.49} & 68.04 & 71.27 \\
    800  & 74.01 & 68.47 & 71.24 \\
    1600 & 74.19 & 68.50 & 71.35 \\
    \bottomrule
    \end{tabular}}
    \caption{
    Effect of compression-free warmup steps.
    }
    \vspace{-6pt}
    \label{tab:warmup_steps}
    \end{table}

    \subsubsection{Effect of Prefix Construction Strategy}
    \label{sec:prefix_construction}
    We compare several query-independent strategies for selecting compressed document representations, including First-$K$, Uniform-$K$, Random-$K$, and attention-based Top-$K$. All variants use the same representation budget and reranking framework, differing only in the selection strategy. As shown in Table~\ref{tab:prefix_strategy}, attention-based Top-$K$ achieves the best overall performance without introducing additional trainable parameters. We therefore adopt attention-based selection as the default document representation selection.
     \begin{table}[htbp]
    \centering
    \renewcommand{\arraystretch}{1.1}
    \resizebox{\linewidth}{!}{
    \begin{tabular}{lccc}
    \toprule
    \textbf{Prefix Construction Strategy} & \textbf{DL19} & \textbf{DL20} & \textbf{Avg.} \\
    \midrule
    First-$K$ tokens & 74.01 & 66.85 & 70.43 \\
    Uniform-$K$ tokens & 74.25 & 67.65 & 70.95 \\
    Random-$K$ tokens & 73.34 & 67.75 & 70.55 \\
    \rowcolor[HTML]{F2F3F5}
    \textbf{Attention-guided Top-$K$} & \textbf{74.30} & \textbf{68.58} & \textbf{71.44} \\
    \bottomrule
    \end{tabular}}
    \caption{
    Effect of different query-independent strategies for selecting document representations. 
    }
    \vspace{-6pt}
    \label{tab:prefix_strategy}
    \end{table}

\section{Conclusion}
In this paper, we introduce \textbf{DoPR}, a compressed document prefix framework for efficient pointwise LLM reranking. DoPR moves reusable document-side computation offline by selecting query-independent document representations and converting them into compressed document prefix states, which are reused during online reranking. Across TREC DL, BEIR, and BRIGHT, DoPR retains 97.1\%--99.5\% of the average NDCG@10 of matched full-document rerankers, while reducing online document-side memory by 8.0$\times$ and achieving up to 8.04$\times$ latency speedup.
These findings suggest that document-side computation in pointwise LLM reranking can be amortized through reusable compressed prefix states.
\newpage
\section*{Limitations}

This work has two main limitations. First, DoPR is most beneficial when documents are reused across many queries, since its efficiency comes from both document-side compression and offline prefix reuse. For rapidly changing collections or scenarios where documents are rarely retrieved repeatedly, the cross-query reuse benefit becomes limited; in such cases, DoPR mainly benefits from the reduced online document budget brought by compression. Second, DoPR shifts part of the computation and storage cost to the offline stage. Although this reduces online reranking latency, large-scale deployment still requires preprocessing the document collection and storing the resulting compressed prefixes, which may introduce additional indexing and storage overhead.

\section*{Ethical considerations}
DoPR aims to improve the efficiency of LLM-based reranking by reducing repeated document-side computation. It does not introduce new data collection procedures or require additional user information beyond standard retrieval inputs. However, like other reranking models, DoPR may inherit biases from the backbone LLM and training data, which can affect the visibility of retrieved documents across domains or user groups. In addition, the offline storage of compressed document prefixes should be managed with the same access control and privacy protections as the original document collection, especially when documents contain sensitive or proprietary content. We recommend careful auditing before deployment in high-stakes retrieval scenarios.
\bibliography{custom}
\newpage
\appendix

% \section{Statement of Related Submission}

% We disclose a concurrent submission by an overlapping set of authors: Submission ID \textbf{420}, \textit{CoT-Rank: Chain-of-Thought Reasoning for Ranking in Retrieval-Augmented Generation}.

% The present work proposes DoPR, which focuses on \textbf{efficient pointwise LLM reranking} by precomputing and reusing query-independent document prefixes. The concurrent CoT-Rank submission studies a different problem: \textbf{improving reranking effectiveness} by distilling chain-of-thought supervision into latent memory slots. Thus, DoPR is centered on cross-query document-side reuse and online efficiency, while CoT-Rank is centered on reasoning-aware supervision for ranking effectiveness. The two submissions have distinct methods and non-overlapping contributions.

\section{Appendix}
% \subsection{Training Curve}
% To better understand the optimization behavior of DoPR, we compare its training loss curve with that of the 8B baseline. As shown in Figure~\ref{fig:training_curve}, although both models gradually converge, DoPR exhibits a more stable training trajectory and achieves lower loss than the baseline over most of the training process, especially after the initial warm-up stage. In contrast, the baseline shows larger fluctuations in the early phase and remains at a higher loss level later in training. These results suggest that DoPR not only improves the final optimization quality, but also leads to a more stable and efficient training process.

\subsection{Training Dynamics with Different Compression Budgets}
\label{sec:app-training-loss-vary-k}

\begin{figure}[hbtp]
    \centering
    \includegraphics[width=\linewidth]{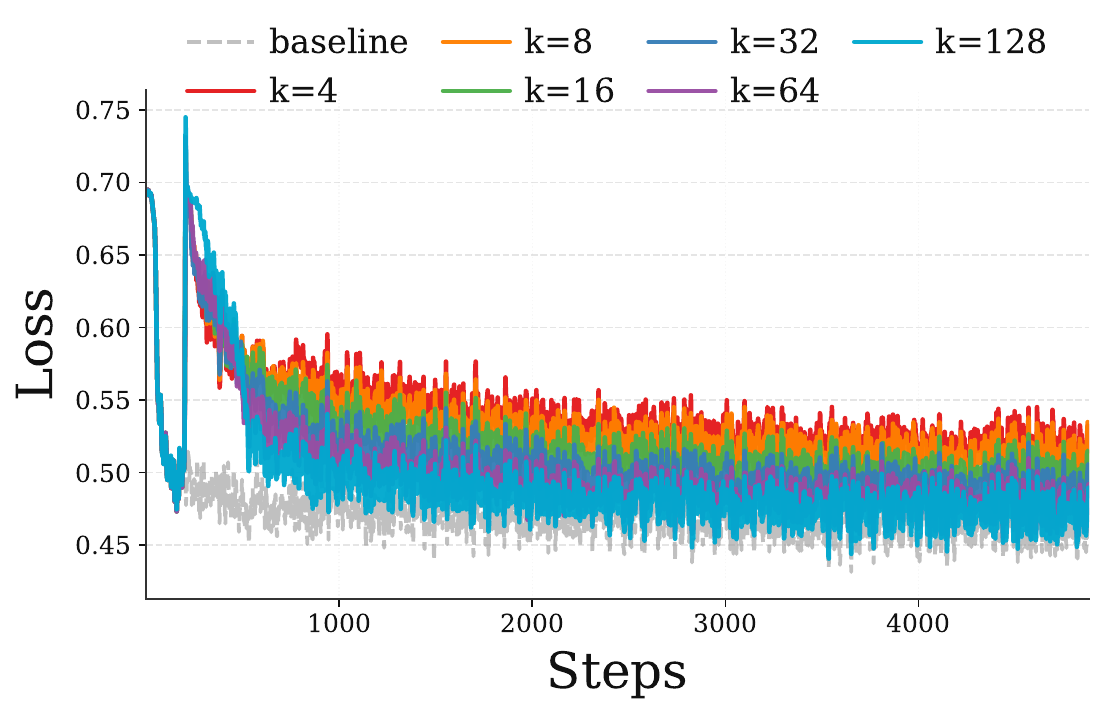}
    \caption{Training loss curves of the 0.6B model under different compression budgets $K_{\text{train}}$.
    % Here, \(K_{\text{train}}\) denotes the number of compressed tokens. 
    % The curves show stable optimization behavior across different compression budgets.
    }
    \label{fig:training-loss-vary-k}
\end{figure}

We further analyze the optimization behavior of our compressed model under different compression budgets. Specifically, we vary \(K_{\text{train}}\), the number of compressed document prefixes, and plot the training loss curves for the 0.6B model. All curves are obtained from the same training setup, with evaluation metrics omitted to focus solely on optimization dynamics. As shown in Figure~\ref{fig:training-loss-vary-k}, the training loss decreases consistently across different values of $K_{\text{train}}$, suggesting that the proposed training procedure remains stable under a broad range of compression budgets.

% \subsection{Effect of Inference Prefix Budget}
% \label{sec:prefix_budget_tradeoff}

\subsection{Additional Experimental Settings}
\label{sec:implementation_details}

We provide additional implementation details that are omitted from the main text due to space constraints in Table~\ref{tab:implementation_details}. These settings are shared by DoPR and the matched Qwen3-Rerank baselines unless otherwise specified. For DoPR, the training prefix budget is fixed to $K_{\text{train}}=32$ by default, while the inference prefix budget $K_{\text{infer}}$ is set according to the dataset-specific document length budget in Table~\ref{tab:token_compression}. The selected document representations are inserted as prefix inputs to the reranker, and the resulting prefix states are stored as multi-layer cached key-value states.

\begin{table*}[htbp]
\centering
\resizebox{0.8\linewidth}{!}{
\begin{tabular}{ll}
\toprule
\textbf{Item} & \textbf{Setting} \\
\midrule
Backbone models & Qwen3-0.6B / Qwen3-4B / Qwen3-8B \\
Optimizer & AdamW \\
Learning rate & $5\times10^{-6}$ \\
Total batch size & 32 \\
Training hardware & 8 NVIDIA A100 GPUs \\
Default training prefix budget & $K_{\text{train}}=32$ \\
Compression-free warmup & 200 steps \\
Inference prefix budget & Dataset-specific $K_{\text{infer}}$ \\
Compression ratio & 8.0$\times$ \\
Inference precision & {bf16} \\
\bottomrule
\end{tabular}}
\caption{
Implementation details and hyperparameter settings used in our experiments.
}
\label{tab:implementation_details}
\end{table*}

\subsection{Dataset-specific Compression Settings}
\label{sec:compression_settings}

Table~\ref{tab:token_compression} summarizes the dataset-specific compression configurations used in our experiments. Since document length budgets vary across benchmarks, we set the retained prefix length proportionally to each dataset's document budget rather than using a single global prefix length. Across all datasets, DoPR uses a fixed 8.0$\times$ compression ratio, retaining 12.5\% of the original document tokens as reusable compressed document prefixes. This proportional setting keeps the compression strength consistent across tasks while allowing the absolute prefix budget to adapt to dataset-specific input lengths, enabling a fair evaluation of online reranking efficiency and effectiveness under heterogeneous benchmark settings.

\begin{table*}[htbp]
    \centering
    \renewcommand{\arraystretch}{1.12}
    \begin{tabular}{ll|ccc}
        \toprule
        \multicolumn{2}{c}{Dataset} & Document Length  & Prefix Length & Compress Ratio \\
        \midrule
        \multirow{2}{*}{TREC DL} & DL 19 & 256 & 32 & 8.0$\times$ \\
         & DL 20 & 256 & 32 & 8.0$\times$ \\
        \midrule
        \multirow{8}{*}{BEIR} & Covid & 1024 & 128 & 8.0$\times$ \\
         & NFCorpus & 1024 & 128 & 8.0$\times$ \\
         & Touche & 2048 & 256 & 8.0$\times$ \\
         & DBPedia & 256 & 32 & 8.0$\times$ \\
         & SciFact & 1024 & 128 & 8.0$\times$ \\
         & Signal & 256 & 32 & 8.0$\times$ \\
         & News & 2048 & 256 & 8.0$\times$ \\
         & Robust & 2048 & 256 & 8.0$\times$ \\
        \midrule
        \multirow{12}{*}{BRIGHT} & Biology & 512 & 64 & 8.0$\times$ \\
         & Economics & 1024 & 128 & 8.0$\times$ \\
         & Earth-Science & 1024 & 128 & 8.0$\times$ \\
         & Psychology & 1024 & 128 & 8.0$\times$ \\
         & Robotics & 1024 & 128 & 8.0$\times$ \\
         & Stackoverflow & 1024 & 128 & 8.0$\times$ \\
         & Sustainable-living & 1024 & 128 & 8.0$\times$ \\
         & Pony & 512 & 64 & 8.0$\times$ \\
         & Leetcode & 1024 & 128 & 8.0$\times$ \\
         & Aops & 1024 & 128 & 8.0$\times$ \\
         & Theoremqa-questions & 1024 & 128 & 8.0$\times$ \\
         & Theoremqa-theorems & 1024 & 128 & 8.0$\times$ \\
        \bottomrule
    \end{tabular}
    \caption{Dataset-specific document lengths and retained prefix lengths used in our experiments.}
    \label{tab:token_compression}
\end{table*}

\subsection{Efficiency Across Datasets and Model Scales}
\label{sec:datasets_efficiency}
Table~\ref{tab:dopr_efficiency_qwen_all} reports additional efficiency results for the full-document Qwen3-Rerank baseline and DoPR across datasets and model scales.
For the Qwen3-Rerank baseline, online inference uses the original document under a dataset-specific document-length budget, whereas DoPR replaces the document with compressed prefixes.
``Memory'' reports the average document-side footprint per query-document pair during online inference, and ``Latency'' reports the corresponding average per-document online inference time.
Therefore, the ``Doc Budget'' column should be interpreted as the configured online document length budget, while memory and latency reflect the realized average online cost.
Across datasets and model scales, DoPR consistently reduces both cache memory and inference latency, with substantially larger relative gains on datasets with longer documents.

\begin{table*}[htbp]
\centering
\renewcommand{\arraystretch}{1.42}
\resizebox{\textwidth}{!}{
\begin{tabular}{llccc ccc c}
\toprule
& & \multicolumn{3}{c}{Qwen3-Rerank} & \multicolumn{3}{c}{DoPR} & \\
\cmidrule(lr){3-5} \cmidrule(lr){6-8}
Model & Dataset & \makecell{Doc \\Budget} & Memory (MiB) & Latency (ms) & \makecell{Doc \\Budget} & Memory (MiB) & Latency (ms) & Speedup \\
\midrule
\multirow{10}{*}{Qwen3-0.6B}
& DL19     & 256  & 28.00  & 17.612  & 32  & 3.50  & 14.600 & 1.21x \\
& DL20     & 256  & 28.00  & 16.055  & 32  & 3.50  & 14.159 & 1.13x \\
& Covid    & 1024 & 112.00 & 41.007  & 128 & 14.00 & 14.372 & 2.85x \\
& NFCorpus & 1024 & 112.00 & 41.238  & 128 & 14.00 & 14.746 & 2.80x \\
& Touche   & 2048 & 224.00 & 101.248 & 64  & 7.00  & 14.673 & 6.90x \\
& DBPedia  & 192  & 21.00  & 15.790  & 32  & 3.50  & 14.285 & 1.11x \\
& SciFact  & 1024 & 112.00 & 40.882  & 128 & 14.00 & 14.384 & 2.84x \\
& Signal   & 256  & 28.00  & 16.165  & 32  & 3.50  & 14.192 & 1.14x \\
& News     & 2048 & 224.00 & 102.148 & 256 & 28.00 & 15.141 & 6.75x \\
& Robust   & 2048 & 224.00 & 101.960 & 256 & 28.00 & 14.268 & 7.15x \\
\midrule
\multirow{10}{*}{Qwen3-4B}
& DL19     & 256  & 36.00  & 29.386  & 32  & 4.50  & 18.430 & 1.59x \\
& DL20     & 256  & 36.00  & 29.271  & 32  & 4.50  & 18.682 & 1.57x \\
& Covid    & 1024 & 144.00 & 121.194 & 128 & 18.00 & 18.095 & 6.70x \\
& NFCorpus & 1024 & 144.00 & 121.204 & 128 & 18.00 & 19.063 & 6.36x \\
& Touche   & 2048 & 288.00 & 293.085 & 64  & 9.00  & 18.330 & 15.99x \\
& DBPedia  & 192  & 27.00  & 23.053  & 32  & 4.50  & 18.040 & 1.28x \\
& SciFact  & 1024 & 144.00 & 121.196 & 128 & 18.00 & 18.406 & 6.58x \\
& Signal   & 256  & 36.00  & 29.282  & 32  & 4.50  & 18.182 & 1.61x \\
& News     & 2048 & 288.00 & 295.088 & 256 & 36.00 & 18.852 & 15.65x \\
& Robust   & 2048 & 288.00 & 294.710 & 256 & 36.00 & 18.299 & 16.11x \\
\midrule
\multirow{10}{*}{Qwen3-8B}
& DL19     & 256  & 36.00  & 37.756  & 32  & 4.50  & 20.547 & 1.84x \\
& DL20     & 256  & 36.00  & 37.599  & 32  & 4.50  & 19.006 & 1.98x \\
& Covid    & 1024 & 144.00 & 150.708 & 128 & 18.00 & 18.743 & 8.04x \\
& NFCorpus & 1024 & 144.00 & 150.361 & 128 & 18.00 & 18.117 & 8.30x \\
& Touche   & 2048 & 288.00 & 355.031 & 64  & 9.00  & 19.405 & 18.30x \\
& DBPedia  & 192  & 27.00  & 34.159  & 32  & 4.50  & 18.829 & 1.81x \\
& SciFact  & 1024 & 144.00 & 150.442 & 128 & 18.00 & 18.822 & 7.99x \\
& Signal   & 256  & 36.00  & 37.582  & 32  & 4.50  & 18.495 & 2.03x \\
& News     & 2048 & 288.00 & 359.434 & 256 & 36.00 & 18.961 & 18.96x \\
& Robust   & 2048 & 288.00 & 358.241 & 256 & 36.00 & 19.070 & 18.79x \\
\bottomrule
\end{tabular}
}
\caption{Efficiency comparison between the baseline Qwen3-Rerank and DoPR across TERC DL and BEIR datasets under Qwen3-0.6B, Qwen3-4B, and Qwen3-8B.}
\label{tab:dopr_efficiency_qwen_all}
\end{table*}

% Rebuttal additions for the appendix.
% Insert after the current subsection "Efficiency Across Datasets and Model Scales".
% The following rebuttal clarifications should instead be incorporated into the main text:
% (1) test-query-independent terminology;
% (2) trainable components and training/offline/online stages;
% (3) the BM25 top-100 -> pointwise DoPR pipeline; and
% (4) the structured mask as a necessary constraint for offline reuse.

\subsection{Generalization Across Model Families}
\label{app:cross-family}

To examine whether DoPR depends on the Qwen architecture, we further evaluate it with Llama-3.2-1B~\citep{grattafiori2024llama}. The full-document reranker and DoPR use the same training data, candidate sets, and evaluation protocol. As shown in Table~\ref{tab:llama-cross-family}, DoPR retains 97.48\% of the full Llama reranker's average NDCG@10 across TREC DL and BEIR benchmarks. Together with the Qwen3 results, this demonstrates that DoPR is not specific to the Qwen architecture and remains effective across both decoder-only model families.

\begin{table*}[htbp]
\centering
\renewcommand{\arraystretch}{1.32}
\setlength{\tabcolsep}{3.3pt}
\resizebox{\textwidth}{!}{
\begin{tabular}{lccccccccccc}
\toprule
Method & DL19 & DL20 & Covid & NFCorpus & Touche & DBPedia & SciFact & Signal & News & Robust & Avg. \\
\midrule
Llama-3.2-Rerank-1B
& 75.76 & 70.78 & 86.05 & 38.77 & 39.76 & 45.61 & 78.98 & 33.63 & 52.08 & 64.37 & 58.58 \\
DoPR-Llama-3.2-1B
& 74.71 & 70.43 & 83.46 & 38.52 & 35.45 & 46.60 & 77.16 & 32.88 & 52.11 & 59.70 & 57.10 \\
\midrule
\textit{Retention}
& 98.61 & 99.51 & 96.99 & 99.36 & 89.16 & \textbf{102.17} & 97.70 & 97.77 & \textbf{100.06} & 92.75 & 97.48 \\
\bottomrule
\end{tabular}}
\caption{Evaluation of Llama-3.2-Rerank-1B
and DoPR-Llama-3.2-1B on TREC DL and BEIR datasets.}
\label{tab:llama-cross-family}
\end{table*}

\begin{table*}[htbp]
\centering
\renewcommand{\arraystretch}{1.32}
\setlength{\tabcolsep}{2.2pt}
\resizebox{\textwidth}{!}{%
\begin{tabular}{lccccccccccccr}
\toprule
Method & Covid & NFCorpus & Touche & DBPedia & SciFact & Signal & News & Robust & ArguAna & FiQA & SciDocs & C-Fever & Avg. \\
\midrule
Qwen3-Rerank-0.6B
& 85.19 & 39.13 & 41.18 & 45.14 & 76.99 & 33.73 & 52.51 & 63.50
& 33.41 & 40.40 & 19.75 & 27.54 & 46.54 \\
DoPR-0.6B
& 84.78 & 38.20 & 37.86 & 44.26 & 76.31 & 31.78 & 51.96 & 59.56
& 33.49 & 35.06 & 19.10 & 26.33 & 44.89 \\
\midrule
\textit{Retention}
& 99.52 & 97.62 & 91.94 & 98.05 & 99.12 & 94.22 & 98.95 & 93.80
& \textbf{100.24} & 86.78 & 96.71 & 95.61 & 96.46 \\
\bottomrule
\end{tabular}%
}
\caption{Expanded zero-shot evaluation of Qwen3-Rerank-0.6B and DoPR-0.6B
across 12 BEIR datasets.}
\label{tab:expanded-beir}
\end{table*}

\subsection{Expanded Out-of-Domain Evaluation}
\label{app:expanded-ood}

TREC DL is the in-domain evaluation, while BEIR and BRIGHT evaluate
out-of-domain transfer without target-domain adaptation. The eight BEIR
datasets in the main evaluation follow the setting used by prior efficient
rerankers. We additionally evaluate Qwen3-Rerank-0.6B and DoPR-0.6B on
ArguAna, FiQA, SciDocs, and Climate-Fever. Across the 12 evaluated BEIR
subsets, DoPR retains 96.46\% of the matched full-document reranker's average
NDCG@10 (Table~\ref{tab:expanded-beir}).

The document prefixes are test-query-independent: each prefix is constructed
from its document without target-collection queries or collection-level
statistics and can be reused across test queries. This does not imply
distribution independence, since the compression parameters are learned from
source-domain query--document pairs.

\subsection{KV-State Reuse and Last-K Selection}
\label{app:last-k}

We compare DoPR with two Last-$K$ alternatives. All variants are independently retrained using identical training data and the same $K$-state storage budget. Raw Last-$K$ KV reuse directly stores the multi-layer KV states of the final $K$ document tokens. Last-$K$ selection retains DoPR's prefix-construction pipeline but replaces attention-guided selection with the final $K$ document states.

As shown in Table~\ref{tab:last-k-baselines}, Last-$K$ selection consistently outperforms Raw Last-$K$ KV reuse, while attention-guided Top-$K$ achieves the best results on both DL19 and DL20. These results indicate that both prefix construction and attention-guided selection contribute to DoPR's effectiveness.
\begin{table}[htbp]
\centering
\resizebox{0.48\textwidth}{!}{
\begin{tabular}{lccc}
\toprule
Method & DL19 & DL20 & Avg. \\
\midrule
Raw Last-K KV reuse & 72.64 & 65.68 & 69.16 \\
Last-K selection & 73.56 & 67.98 & 70.77 \\
Attention-guided Top-K & \textbf{74.30} & \textbf{68.58} & \textbf{71.44} \\
\bottomrule
\end{tabular}}
\caption{Comparison of Raw Last-$K$ KV reuse, Last-$K$ selection,
and attention-guided Top-$K$ on TREC DL19 and DL20.}
\label{tab:last-k-baselines}
\end{table}
\subsection{Efficiency on BRIGHT}
\label{app:bright-efficiency}

We further evaluate the online efficiency of Qwen3-Rerank-0.6B and DoPR-0.6B across all 12 BRIGHT subsets. DoPR retains 98.6\% of the full-document reranker's average NDCG@10 while reducing the average document-state footprint by 8.0$\times$ and achieving a 4.89$\times$ average latency speedup. The speedup is more pronounced on the 1024-token subsets, ranging from 5.02$\times$ to 5.99$\times$. The average speedup in Table~\ref{tab:bright-efficiency} is computed as the ratio between the mean baseline latency and the mean DoPR latency across subsets.
\begin{table*}[htbp]
\centering
\renewcommand{\arraystretch}{1.42}
\setlength{\tabcolsep}{3.2pt}
\resizebox{\textwidth}{!}{
\begin{tabular}{lccccccc}
\toprule
& \multicolumn{3}{c}{Qwen3-Rerank-0.6B} & \multicolumn{3}{c}{DoPR-0.6B} & \\
\cmidrule(lr){2-4}\cmidrule(lr){5-7}
Dataset & \makecell{Doc \\Budget} & Memory (MiB) & Latency (ms)
& \makecell{Doc \\Budget} & Memory (MiB) & Latency (ms) & Speedup \\
\midrule
Biology             & 512  & 56.00  & 17.267 & 64  & 7.00  & 9.414  & 1.83$\times$ \\
Economics           & 1024 & 112.00 & 56.886 & 128 & 14.00 & 9.814  & 5.80$\times$ \\
Earth Science       & 1024 & 112.00 & 51.457 & 128 & 14.00 & 9.408  & 5.47$\times$ \\
Psychology          & 1024 & 112.00 & 56.839 & 128 & 14.00 & 9.495  & 5.99$\times$ \\
Robotics            & 1024 & 112.00 & 79.146 & 128 & 14.00 & 15.772 & 5.02$\times$ \\
Stack Overflow      & 1024 & 112.00 & 79.522 & 128 & 14.00 & 15.793 & 5.04$\times$ \\
Sustainable Living  & 1024 & 112.00 & 56.688 & 128 & 14.00 & 9.688  & 5.85$\times$ \\
Pony                & 512  & 56.00  & 17.380 & 64  & 7.00  & 9.514  & 1.83$\times$ \\
LeetCode            & 1024 & 112.00 & 79.714 & 128 & 14.00 & 15.772 & 5.05$\times$ \\
AoPS                & 1024 & 112.00 & 51.762 & 128 & 14.00 & 9.250  & 5.60$\times$ \\
Theoremqa-questions & 1024 & 112.00 & 51.825 & 128 & 14.00 & 9.673  & 5.36$\times$ \\
Theoremqa-theorems  & 1024 & 112.00 & 51.828 & 128 & 14.00 & 9.530  & 5.44$\times$ \\
\midrule
Average             & --   & 102.67 & 54.19  & --  & 12.83 & 11.09  & 4.89$\times$ \\
\bottomrule
\end{tabular}}
\caption{Efficiency comparison between the baseline Qwen3-Rerank and DoPR across
the 12 BRIGHT subsets.}
\label{tab:bright-efficiency}
\end{table*}

\subsection{Larger Rerankers Under Comparable Online Budgets}
\label{app:larger-rerankers}

We compare DoPR-8B with the smaller Qwen3-Rerank-0.6B under the same online evaluation protocol. As shown in Table~\ref{tab:larger-reranker-budget},
DoPR-8B achieves slightly higher effectiveness on both datasets while
maintaining comparable or lower latency and a smaller document-state
footprint. This comparison concerns online document processing only.
DoPR-8B still requires substantially more memory for model parameters.

\begin{table*}[ht]
\centering
\resizebox{0.8\textwidth}{!}{
\begin{tabular}{llccc}
\toprule
Dataset & Method & NDCG@10 & Memory (MiB) & Latency (ms) \\
\midrule
DL19 & Qwen3-Rerank-0.6B & 75.30 & 28.0 & \textbf{17.61} \\
DL19 & DoPR-8B & \textbf{76.17} & \textbf{4.5} & 20.55 \\
\midrule
Covid & Qwen3-Rerank-0.6B & 85.19 & 112.0 & 41.01 \\
Covid & DoPR-8B & \textbf{85.42} & \textbf{18.0} & \textbf{18.74} \\
\bottomrule
\end{tabular}}
\caption{Comparison between DoPR-8B and the full-document
Qwen3-Rerank-0.6B.}
\label{tab:larger-reranker-budget}
\end{table*}

\end{document}